\documentclass[11pt,superscriptaddress,aps,prd,preprint]{revtex4}
\usepackage{amsmath}

\makeatletter
\usepackage[T1]{fontenc}
\usepackage{amsmath}
\usepackage{amssymb}
\usepackage{graphicx}
\usepackage{tabularx}

\usepackage{slashed}
\usepackage{xcolor}

\newcommand{\bea}{\begin{eqnarray}}
\newcommand{\eea}{\end{eqnarray}}

\begin{document}

\title{On Scalar Electromagnetism in Phase Space }

\author{R. G. G. Amorim}\email[]{ronniamorim@gmail.com}
 \affiliation{Faculdade Gama, Universidade de Bras\'{i}lia, Setor Leste
(Gama), 72444-240, Bras\'{i}lia-DF, Brazil.}

\author{J. S da Cruz Filho}\email[]{ze_filho@fisica.ufmt.br}
\affiliation{Instituto de F\'{\i}sica, Universidade Federal de Mato Grosso,\\
78060-900, Cuiab\'{a}, Mato Grosso, Brazil}

\author{A. F. Santos}\email[]{alesandroferreira@fisica.ufmt.br}
\affiliation{Instituto de F\'{\i}sica, Universidade Federal de Mato Grosso,\\
78060-900, Cuiab\'{a}, Mato Grosso, Brazil}

\author{S. C. Ulhoa}\email[]{sc.ulhoa@gmail.com}
\affiliation{Instituto de F\'{i}sica,
Universidade de Bras\'{i}lia, 70910-900, Bras\'{i}lia, DF,
Brazil.}

\begin{abstract}
In this paper the interaction of a scalar field and the electromagnetic field in phase space is analyzed. The scattering process is calculated up to first order in the Planck constant which is obtained by an expansion of the Moyal product in phase space. The transition amplitude is calculated in the same context.

\end{abstract}

\maketitle

\section{Introduction}

Classical Mechanics is described in terms of two equivalent formalisms. The first one is established in the configuration space in terms of the Lagrangian function which is constructed out of an invariant quantity under a given symmetry in such a space. The second one is derived from the Hamiltonian function which is defined in a simplectic manifold, also known as the phase space. The main advantage of such a formalism is its inherent physical interpretation once the Hamiltonian is related to the energy of the system. It should be noted that for every mechanics there exists these both formalisms. It holds for Quantum or Classical Mechanics. However the definition of the phase space in Quantum Mechanics is not direct as we desire due to the uncertainty relations arising in the quantum world.

In 1932 Wigner come out with an idea to establish the phase space in Quantum Mechanics for the first time \cite{wig1,wig2}. He proposed a kind of Fourier transformation of the density matrix while studying corrections for Statistical Mechanics. Since then another attempts have been made such as work by Torres-Vegas \cite{torres01,torres02}. We point out that up to our knowledge only Wigner proposal has a well defined classical limit. Such a feature is very important in order to interpret what is going on at the quantum level. Therefore based on Wigner formalism it is possible to map a quantum dynamics into the respective phase space by means the use of the star or Moyal product. Thus the Wigner function has more information about than its quantum counterpart \cite{Kenfack}. For instance it can be used to analyze how classical is the system. Such a powerful method has been recently used to study Schwinger pair production in strong electric fields \cite{Hebenstreit, Hebenstreit1}. Particularly such a formalism has been used to construct the Galilean symmetry in phase space and its consequences, as an example, the Schrödinger equation in phase space is obtained \cite{seb1,amorim1,amorim2,amorim3,amorim00,amorim5,amorim4}. An extension for the relativistic symmetry was accomplished in reference \cite{amorim1} where the Klein-Gordon and Dirac equation in the phase space was studied. The solutions of these phase space equations are called amplitudes which by themselves do not have physical interpretation. However, the Wigner function for a given system can be provided by the Moyal product between this amplitude and its conjugated complex. This calculation provides a new method to calculate the Wigner function. Another advantage of this phase space formalism, due to the fact that amplitudes in phase space are complex quantities, is the possibility of analyzing gauge symmetries in phase space, which is not possible in traditional Wigner formalism \cite{seb1,amorim1}.

In this article the interaction between scalar and electromagnetic fields in phase space is analyzed. On one hand it is not a realistic interaction, hence it can be viewed as a toy-model. On the other hand the interaction of such basic fields can shed light on the structure of the phase space. Such a interaction leads to the so called scalar electrodynamics. It's a model that explains the symmetry breaking by means quantum corrections \cite{scalar01}. In fact scalar electrodynamics has a geometrical realization in a seven-dimensional space \cite{scalar02}. This mechanisms of symmetry breaking has been applied to cosmology in order to explain primordial magnetic fields which supposedly arise during an inflationary period \cite{scalar03}.
Renormalization group methods are employed to study  the behavior of the  ultraviolet completion of  five-dimensional scalar QED \cite{scalar04} and effective potential in massless theories \cite{scalar05}.

This article is divided as follows. In section \ref{2}, we briefly review how to introduce the phase space. In section \ref{3}, we calculate the propagators and vertices related to the fields. In section \ref{4}, we show the transition amplitude of the process of interaction between two scalar fields and a photon. Finally in the last section we revise our results and presents our final points.

\section{Scalar Field in Phase Space} \label{2}

In this section we present a brief outline of the symplectic representation of Poincaré algebra. From this representation we obtain the Klein-Gordon equation in phase space. The relation between symplectic representation and traditional Wigner phase space formalism is also presented.

Poincaré-Lie algebra is given by the set of commutation relations

\begin{equation}\nonumber
[\widehat{M}_{\mu\nu},\widehat{P}_{\sigma}]=i(g_{\nu\sigma}\widehat{P}_{\mu}-g_{
\sigma\mu}\widehat{P}_{\nu}),
\end{equation}

\begin{equation}\nonumber
[\widehat{P}_{\mu},\widehat{P}_{\nu}]=0,
\end{equation}

\begin{equation}\nonumber
[\widehat{M}_{\mu\nu},\widehat{M}_{\sigma\rho}]=-i(g_{\mu\rho}\widehat{M}_{\nu\sigma}-
g_{\nu\rho}\widehat{M}_{\mu\sigma}+g_{\mu\sigma}\widehat{M}_{\rho\nu}-g_{\nu
\sigma}\widehat{M}_{\rho\mu}),
\end{equation}
where $\widehat{M}_{\mu\nu}$ stand
for rotations and $\widehat{P}_{\mu}$ for translations.

This algebra is satisfied by the star-operators defined in phase space by

\begin{equation}\label{op1}
\widehat{P}^{\mu}=p^{\mu}\star=
p^{\mu}-\frac{i}{2}\frac{\partial}{\partial q_{\mu}},
\end{equation}

\begin{equation}\label{op2}
\widehat{Q}^{\mu}=q^{\mu}\star =
q^{\mu}+\frac{i}{2}\frac{\partial}{\partial p_{\mu}},
\end{equation}

\begin{equation}\label{op3}
\widehat{M}_{\nu\sigma}=M_{\nu\sigma}\star
=\widehat{Q}_{\nu}\widehat{P}_{\sigma}-\widehat{Q}_{\sigma}\widehat{P}_{\nu}.
\end{equation}

The operators $\widehat{Q}^{\mu}$, $\widehat{P}^{\mu}$ and $\widehat{M}_{\nu\sigma}$ are defined in the Hilbert space $\mathcal{H}(\Gamma)$, constructed with complex functions in phase space $\Gamma$ represented by $\phi(q,p)$, where  $(q,p)=(q^{\mu },p^{\mu })$ \cite{amorim1}.

The Casimirs of this algebra can be built up from the Pauli-Lubanski
matrices, $\widehat{W}_{\mu }=\frac{1}{2}\varepsilon _{\mu \nu \rho
\sigma }\widehat{M}^{\nu \sigma }\widehat{P}^{\rho }$, where
$\varepsilon _{\mu \nu \rho \sigma }$ is the Levi-Civita symbol. The
invariants are \cite{amorim1}:
\begin{eqnarray}\label{eq1}
\widehat{P}^{2} &=&\widehat{P}^{\mu }\widehat{P}_{\mu },  \\
\widehat{W} &=&\widehat{W}^{\mu }\widehat{W}_{\mu },
\end{eqnarray}
where $\widehat{P}^{2} $ stands for the mass shell condition and
$\widehat{W}$ for the spin.

In the following we use such a
representation to derive equations in phase space for spin 0.
To determine the Klein-Gordon field equation, we
consider a scalar representation in $\mathcal{H}(\Gamma)$. In this
case we can use the invariant $ \widehat{P} ^2 $ given in Eq.(\ref{eq1}) to
write
\begin{eqnarray*}
\widehat{P}^2\phi(q,p)&=&(p^2)\star \phi(q,p) \\
&=&(p^{\mu }\star p_{\mu }\star) \phi(q,p)=m^2\phi(q,p),
\end{eqnarray*}
where $m$ is a constant fixing the representation and interpreted as
mass, such that the mass shell condition is satisfied. Using Eq.(\ref{op1}) we
obtain
\begin{equation}
(p^{\mu }p_{\mu }-ip^{\mu}\frac{\partial }{\partial q^{\mu }}-\frac{1}{4}%
\frac{\partial }{\partial q^{\mu }}\frac{\partial }{\partial q_{\mu }}%
)\phi(q,p)=m^2\phi(q,p) ,
\end{equation}
which is the Klein-Gordon equation in phase space \cite{amorim1}. This equation can be derived from lagrangian given by \cite{amorim3}
\begin{eqnarray}
 \mathcal{L}
&=&- \left( \widehat{P}^{\mu}  \star \phi  \right) \star \left( \phi^{\ast} \star \widehat{P}_{\mu} \right)
+ m^2 \phi^{\ast} \star \phi.
 \label{eq2}
\end{eqnarray}

Consider Klein-Gordon equation for $\phi(q,p)$ and $\phi^{\ast}(q,p)$,
\begin{equation}\label{p1}
p^2\star\phi(q,p)=m^2\phi(q,p),
\end{equation}
\begin{equation}\label{p2}
\phi^{\ast}(q,p)\star p^2=m^2\phi^{\ast}(q,p).
\end{equation}
Multiplying the right-hand side of Eq.(\ref{p1}) by $\phi^{\ast}(q,p)$ and the left-hand side of Eq.(\ref{p2}) by $\phi(q,p)$ and using the associativity of Moyal product we obtain
\begin{equation}\label{p3}
p^2\star(\phi(q,p)\star\phi^{\ast}(q,p))=m^2(\phi(q,p)\star\phi^{\ast}(q,p)),
\end{equation}
\begin{equation}\label{p4}
(\phi(q,p)\star\phi^{\ast}(q,p))\star p^2=m^2(\phi(q,p)\star\phi^{\ast}(q,p)).
\end{equation}
Defining $f(q,p)=\phi(q,p)\star\phi^{\ast}(q,p)$ and subtracting Eq.(\ref{p4}) of Eq.(\ref{p3}), we get
\begin{equation}\label{p5}
p^2\star f(q,p)-f(q,p)\star p^2=0.
\end{equation}
The Moyal product (star product) in phase space is defined by
\begin{equation}\label{p6}
a_{w}(q,p)\star b_{w}(q,p)= a_{w}(q,p)
\exp\left[\frac{i}{2}\left(\frac{\overleftarrow{\partial}}{\partial
q}\frac{\overrightarrow{\partial}}{\partial
p}-\frac{\overleftarrow{\partial}}{\partial
p}\frac{\overrightarrow{\partial}}{\partial q}\right)\right]b_{w}(q,p).
\end{equation}
Using this relation, Eq.(\ref{p5}) can be written by
\begin{equation}\label{p7}
p_{\mu}\frac{\partial f(q,p)}{\partial q^{\mu}}=0.
\end{equation}
Eq.(\ref{p7}) is a well know equation to relativistic Wigner function. In this way, $f(q,p)$ is the relativistic Wigner function $f_W(q,p)$.

In this sense, the association of this representation with the  Wigner formalism is
given by \cite{seb1, amorim1}
\begin{equation}\label{eq3}
f_{W}(q,p)=\phi(q,p)\ast\phi^{\dagger}(q,p),
\end{equation}
where $f_{W}(q,p)$ is the relativistic Wigner function. The properties of Wigner function can be easily derived. It is useful emphasize that the phase space amplitudes $\phi(q,p)$ do not have physical interpretation by themselves, but the Wigner function can be calculated by Eq.(\ref{eq3}) and provide the physical interpretation for this formalism.

\section{Lagrangian in phase space} \label{3}
\hspace{0,6cm}
The lagrangian that describes the interaction between the  klein-Gordon and electromagnetic fields in phase space is  given by
\begin{eqnarray}
 \mathcal{L}_{total}
&=&- \left( D^{\mu}  \star \phi  \right) \star \left( \phi^{\ast} \star D_{\mu} \right)
+ m^2 \phi^{\ast} \star \phi -\frac{1}{4} \mathcal{F}_{\mu \nu}\mathcal{F}^{\mu \nu}.
 \label{eq:02}
\end{eqnarray}
where $\phi= \phi(q,p)$, $\phi^{\ast}=  \phi^{\ast}(q',p')$ and the covariant derivative is $ D^{\mu} \star = p^{\mu} \star +  i e A^{\mu} \star $. It should be noted that the electromagnetic part of the Lagrangian does not have the star product, this is because such a product trivializes for two factors. In addition the operator $D^\mu\star$ is constructed in order to leave the Lagrangian invariant under U1 group transformations. As a matter of fact defining gauge symmetry in phase space is quite a difficult task, thus in this article such a feature will not be detailed. In fact we can explore how the scalar field interacts with electromagnetism by expanding the Moyal product up to first order. Hence the electromagnetic sector remains unaltered and corrections, concerning scalar electrodynamics, can be provided. It should be pointed out that on one hand the propagator of scalar field is already known from previous works, on another hand its interaction with electromagnetic field has never been calculated. We chose to perform such a calculation approximately in two flanks, in the Moyal product and in the calculation of the processes themselves. Expanding the covariant derivative we obtain
\begin{eqnarray}
 \mathcal{L}_{total}&=&-\left( p^\mu \star \phi  \right) \star \left( \phi^{\ast} \star p_\mu \right)
 + i\left(p_\mu \star \phi \star \phi^{\ast}\star A^\mu
 + A_\mu  \star \phi \star \phi^{\ast}\star p^\mu   \right) \notag\\
 &-& A_\mu  \star \phi \star \phi^{\ast}\star A^\mu
+  m^2  \phi^{\ast} \star \phi  -\frac{1}{4} \mathcal{F}_{\mu \nu}\mathcal{F}^{\mu \nu}. 
 \label{eq:03}
\end{eqnarray}
Then the interaction part is
\begin{eqnarray}
\mathcal{L}_{int}&=&
  i\left(p_\mu \star \phi \star \phi^{\ast}\star A^\mu
 + A_\mu  \star \phi \star \phi^{\ast}\star p^\mu   \right) -
  A_\mu  \star \phi \star \phi^{\ast}\star A^\mu ,\nonumber\\
&=&i\left( (\widehat{p}_{\mu}) ( \phi \star \phi^{\ast}) (\widehat{A}^\mu)^{\ast}\right)
  +
  i\left( (\widehat{A}_\mu)( \phi \star \phi^{\ast})(\widehat{p}^{\mu})^{\ast} \right)
 - (\widehat{A}_\mu)( \phi \star \phi^{\ast})(\widehat{A}^{\mu})^{\ast}
\end{eqnarray}
where $\widehat{p}^{\mu} =  {p}^{\mu} \star $ and $ \widehat{A}_\mu = A_\mu \star $ have been used.


Now let's write explicitly the propagators of the Klein-Gordon and electromagnetic field and the interaction term.

\subsection{Klein-Gordon Propagator}

Here the free lagrangian of Klein-Gordon field is considered to construct the propagator. This lagrangian in phase space is
\begin{equation}
 \mathcal{L}_{\phi}=-\left( p^\mu \star \phi  \right)\star \left( \phi^{\ast} \star p_\mu \right)
+  m^2  \phi^{\ast} \star \phi.
 \label{eq:01}
\end{equation}
Using
\begin{equation}
 {p}^{\mu} \star=\widehat{p}^{\mu} = p^\mu - \frac{i}{2}\frac{\partial }{\partial q_\mu}.\label{defp}
 \end{equation}
 the lagrangian becomes
\begin{eqnarray}
 \mathcal{L}_{\phi}
 &=&-\frac{1}{4}\frac{\partial  \phi }{ \partial q_\mu}
 \frac{\partial  \phi^{\ast}}{ \partial q^\mu}
+\frac{1}{2}i p^\mu
\left(\phi^{\ast}\frac{\partial\phi }{\partial q^\mu}
-\phi \frac{\partial \phi^{\ast}}{\partial q^\mu}
 \right)-
(p^\mu p_\mu-m^2)\phi \phi^{\ast}.
\end{eqnarray}
Rewritten the quantities
\begin{equation*}
 \phi  \frac{\partial  \phi^{\ast}}{ \partial q^\mu}=
  \frac{\partial  }{ \partial q^\mu}(\phi\phi^{\ast})
  - \phi^{\ast}\frac{\partial  \phi}{ \partial q^\mu}
\end{equation*}
and
\begin{equation}
\frac{\partial  \phi }{ \partial q_\mu}
 \frac{\partial  \phi^{\ast}}{ \partial q^\mu}=
  \frac{\partial   }{ \partial q_\mu} \left(
 \frac{\partial  \phi}{ \partial q^\mu} \phi^{\ast}\right)
 -\phi^{\ast} \frac{\partial^2  \phi}{ \partial q^\mu \partial q_\mu},
\end{equation}
the lagrangian is written as
\begin{eqnarray}
 \mathcal{L}_{\phi}
 &=& \frac{1}{4} \phi^{\ast} \frac{\partial^2  \phi}{ \partial q^\mu \partial q_\mu}
+i  \phi^{\ast}p^\mu \frac{\partial  \phi}{ \partial q^\mu }
-\phi^{\ast}(p^\mu p_\mu-m^2)\phi, \notag
\end{eqnarray}
where the total divergence terms vanishes. In terms of the propagator we obtain
\bea
 \mathcal{L}_{\phi}=\phi^{\ast} \mathcal{D}^{-1}_{\phi}\phi
\eea
where
\begin{eqnarray}
 \mathcal{D}^{-1}_{\phi}
 &=& \frac{1}{4}  \frac{\partial^2 }{ \partial q^\mu \partial q_\mu}
+i p^\mu \frac{\partial  }{ \partial q^\mu }
-p^2-m^2
\end{eqnarray}
is the Klein-Gordon propagator that is represented in FIG.1. It should be noted that the scalar field propagator above contains the usual form which is derived in classical field plus a combination of derivatives with respect to the coordinates which is exclusively a prediction of phase space.
\begin{figure}[h]
\includegraphics[scale=0.9]{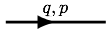}
\caption{The Klein-Gordon propagator or scalar field propagator in phase space. It propagates
both $\phi$ and $\phi^*$.}
\end{figure}

\subsection{Photon Propagator}

The electromagnetic lagrangian in phase space is given by
\bea
\mathcal{L}_{A}=-\frac{1}{4} \mathcal{F}^{\mu\nu}\mathcal{F}_{\mu\nu}
\eea
where
\begin{equation}
\mathcal{F}^{\mu\nu} =   \frac{\partial A^{\mu}}{\partial q_{\nu}}
-\frac{\partial A^{\nu}}{\partial q_{\mu}} + \left\{ A_{\mu}, A_{\nu} \right\}
\end{equation}
using that $\left\{a,b\right\}=a\star b - b \star a $, we obtain $\left\{ A_{\mu}, A_{\nu} \right\}=0$. Considering that $ A_{\mu}\equiv  A_{\mu}(q)$ we can write
\begin{eqnarray}
\mathcal{F}^{\mu\nu} \mathcal{F}_{\mu\nu}& = &  \left(\frac{\partial A^{\mu}}{\partial q_{\nu}}
-\frac{\partial A^{\nu}}{\partial q_{\mu}} \right)
\left(\frac{\partial A_{\mu}}{\partial q^{\nu}}
-\frac{\partial A_{\nu}}{\partial q^{\mu}} \right) \notag\\
& = &
\frac{\partial A^{\mu}}{\partial q_{\nu}}\frac{\partial A_{\mu}}{\partial q^{\nu}}
-\frac{\partial A^{\mu}}{\partial q_{\nu}}\frac{\partial A_{\nu}}{\partial q^{\mu}}
-\frac{\partial A^{\nu}}{\partial q_{\mu}}\frac{\partial A_{\mu}}{\partial q^{\nu}}
+\frac{\partial A^{\nu}}{\partial q_{\mu}}\frac{\partial A_{\nu}}{\partial q^{\mu}} \notag\\
&=& -A^{\mu} \frac{\partial^2 }{\partial q_{\nu} \partial q^{\nu}} A_{\mu}
+
A^{\mu} \frac{\partial^2 }{\partial q_{\nu} \partial q^{\mu}} A_{\nu}
+
A^{\nu} \frac{\partial^2 }{\partial q_{\mu} \partial q^{\nu}} A_{\mu}
-
A^{\nu} \frac{\partial^2 }{\partial q_{\mu} \partial q^{\nu}} A_{\nu}
\end{eqnarray}
Then
\bea
\mathcal{L}_{A}=-\frac{1}{4} \Biggl[ -A^{\mu} \frac{\partial^2 }{\partial q_{\nu} \partial q^{\nu}} A_{\mu}
+
A^{\mu} \frac{\partial^2 }{\partial q_{\nu} \partial q^{\mu}} A_{\nu}
+
A^{\nu} \frac{\partial^2 }{\partial q_{\mu} \partial q^{\nu}} A_{\mu}
-
A^{\nu} \frac{\partial^2 }{\partial q_{\mu} \partial q^{\mu}} A_{\nu}\Biggl]
\eea
using the Lorentz gauge, i.e.,  $ \frac{\partial A_{\nu}}{\partial q_{\nu}}=0$, the lagrangian becomes
\begin{eqnarray}
\mathcal{L}_{A}& = & \frac{1}{2} A^{\mu} \frac{\partial^2 }{\partial q_{\nu} \partial q^{\nu}} A_{\mu}=A^{\mu} \mathcal{G}^{-1}_{\mu \lambda} A_{\lambda}.
\end{eqnarray}
Then the photon propagator is defined as
\begin{eqnarray}
\mathcal{G}^{-1}_{\mu \lambda}& = & \frac{1}{2}
g_{\mu \lambda}
 \frac{\partial^2 }{\partial q_{\nu} \partial q^{\nu}},\label{PP}
\end{eqnarray}
which is represented in FIG.2.
\begin{figure}[h]
\includegraphics[scale=0.9]{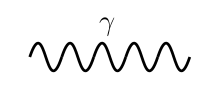}
\caption{ The photon propagator in phase space.}
\end{figure}

\subsection{Interaction term}

Now let's to work with the interaction part that is given by
\begin{eqnarray}
\mathcal{L}_{int}&=&\mathcal{L}_{int}^1+\mathcal{L}_{int}^2
\end{eqnarray}
where
\bea
\mathcal{L}_{int}^1&=&  i\left(p_\mu \star \phi \star \phi^{\ast}\star A^\mu
 + A_\mu  \star \phi \star \phi^{\ast}\star p^\mu   \right)\label{I1}\\
 \mathcal{L}_{int}^2&=&-
  A_\mu  \star \phi \star \phi^{\ast}\star A^\mu,\label{I2}
\eea
with $ A^{\mu}\equiv  A^{\mu}(q^\mu)$.  Using (\ref{defp}) the first term, eq. (\ref{I1}), is written as
\bea
\mathcal{L}_{int}^1&=&i\Biggl[\left(  p_\mu - \frac{i\hbar}{2}\frac{\partial  }{ \partial q^\mu } \right)
 (\phi \star \phi^{\ast})A^\mu \left(  q_\mu - \frac{i\hbar}{2}\frac{\overleftarrow{\partial } }{ \partial p^\mu } \right)\nonumber\\
 &+&A_\mu \left( q^\mu + \frac{i\hbar}{2}\frac{\partial  }{ \partial p^\mu } \right)
(\phi \star \phi^{\ast}) \left(  p^\mu + \frac{i\hbar}{2}\frac{\overleftarrow{\partial}  }{ \partial q_\mu } \right)\Biggl].
\eea
Using the  identity
\begin{eqnarray}
p_k (f \star g) &=& f \star (p_k g) -\frac{i}{2} (\partial_{qk}f) \star g  \notag \\
&=&( p_k f) \star g +\frac{i}{2} f \star  (\partial_{qk}g),
\end{eqnarray}
we get
\bea
\mathcal{L}_{int}^1&=&i\Biggl\{\left[\phi \star ( p_\mu \phi^{\ast})-\left(\frac{i\hbar}{2} \frac{\partial  \phi }{ \partial q^\mu }\right) \star \phi^{\ast}
   \right]A^\mu \left(  q_\mu - \frac{i\hbar}{2}\frac{\overleftarrow{\partial } }{ \partial p^\mu } \right)\Biggl]\nonumber\\
   &+&A_\mu \left( q^\mu + \frac{i\hbar}{2}\frac{\partial  }{ \partial p^\mu } \right)\left[ ( \phi p^\mu ) \star \phi^{\ast}+\left(\frac{i\hbar}{2} \frac{\partial  \phi }{ \partial q^\mu }\right) \star \phi^{\ast} \right\}\nonumber\\
   &=& i\left(  \phi \star ( p_\mu \phi^{\ast} )+( \phi p^\mu ) \star \phi^{\ast} \right) A_\mu
\eea
Using the definition of Moyal product
\begin{eqnarray}
A \star B  = A  \exp
\left[ \frac{i\hbar }{2}\left( \frac{\overleftarrow{\partial }}{\partial q_\mu}%
\frac{\overrightarrow{\partial }}{\partial p^\mu}-\frac{\overleftarrow{\partial }%
}{\partial p^\mu}\frac{\overrightarrow{\partial }}{\partial q_\mu}\right) \right]
B,
\end{eqnarray}
and considering the limit $\hbar \ll 1$,
\begin{eqnarray}
A \star B  &=& AB + \frac{i\hbar }{2}
\left( \frac{\partial A }{\partial q_\mu}%
\frac{ \partial B }{\partial p^\mu}-\frac{\partial A }%
{\partial p^\mu}\frac{\partial B }{\partial q_\mu}\right).\label{hp}
\end{eqnarray}
Then
\begin{eqnarray}
 \mathcal{L}_{int}^1&=&  A_\mu  \phi^{\ast}  \bigg[ 2i   p^\mu  -   \frac{ \hbar }{2}  \frac{\overrightarrow{\partial}  }{\partial q_\mu}
  +
 \frac{ \hbar }{2}  \frac{\overleftarrow{\partial}  }{\partial q'_\mu}
 - \hbar  \frac{\overleftarrow{\partial }}{\partial p'^\nu}  p^\mu  \frac{ \overrightarrow{\partial}  }{\partial q_\nu}
 + \hbar  \frac{\overleftarrow{\partial } }{\partial q'_\nu}  p^\mu  \frac{ \overrightarrow{\partial} }{\partial p^\nu}
  \bigg] \phi,
\end{eqnarray}
which may be rewritten as
\begin{eqnarray}
 \mathcal{L}_{int}
 &=&  A_\mu  \phi^{\ast}(q',p') \,\left[ \mathcal{V}^{\mu}_{\phi \phi^{\ast}A}\right]\,\phi (q,p)
\end{eqnarray}
where the vertex is defined as
\bea
 \mathcal{V}^{\mu}_{\phi \phi^{\ast}A}
 &=&  2i   p^\mu
 + \hbar \left[ \frac{1}{2} \left(\frac{\partial}{\partial q'_\mu}
 - \frac{\partial}{\partial q_\mu}\right)
  + p^\mu  \left( \frac{\partial^2}{\partial q'_\nu \partial p^\nu}
 - \frac{\partial^2}{\partial p'^\nu \partial q_\nu}\right) \right].\label{TV}
\eea
This vertex is represented in FIG.3. We point out that the vertex of such a process has a correction of first order in terms of $\hbar$ when compared to QED.
\begin{figure}[h]
\includegraphics[scale=0.8]{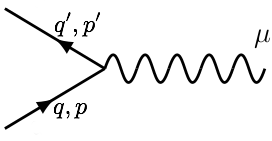}
\caption{ Three-point vertex in phase space, i.e., scalar-photon interaction ($\phi\phi^*A$), with (p,q,) and (p', q') being the 4-momentum and 4-position of scalar fields in phase space.}
\end{figure}

In this theory there is another vertex, 4-point vertex, that is given by the lagrangian
\begin{eqnarray}
 \mathcal{L}_{int}^2
 &=& -  A_\mu ( \phi \star \phi^{\ast}) A^\mu  \notag\\
  &=& -  A_\mu  A^\mu \left(   \phi \phi^{\ast}
 +
   \frac{i \hbar }{2}   \frac{\partial \phi}{\partial q_\nu} \frac{\partial \phi^{\ast}}{\partial p^\nu}
   - \frac{i \hbar }{2}   \frac{\partial \phi}{\partial p^\nu} \frac{\partial \phi^{\ast}}{\partial q_\nu} \right)\nonumber\\
   &=&-  A_\mu  A^\mu    \phi^{\ast}
  \left[ \mathcal{V}_{\phi \phi^{\ast}AA}\right] \phi
\end{eqnarray}
where the eq. (\ref{hp}) has been used and
\begin{eqnarray}
 \mathcal{V}_{\phi \phi^{\ast}AA}
 &=&1+ \frac{i \hbar }{2}  \left( \frac{\partial^2}{\partial q_\nu \partial p'^\nu}
   -    \frac{\partial^2 }{\partial p^\nu \partial q'_\nu} \right).
\end{eqnarray}
It is the vertex of the quartic interaction that is represented in FIG.4. Similar to the expression (41), here there is a first order correction of $\hbar$.
\begin{figure}[h]
\includegraphics[scale=0.8]{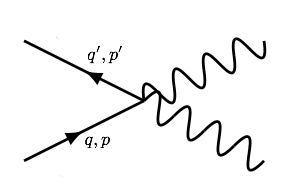}
\caption{ Four-point vertex ($\phi \phi^{\ast}AA$) in phase space. Note that, (p,q,) and (p', q') being the 4-momentum and 4-position of scalar fields in phase space.}
\end{figure}

\section{Transition amplitude} \label{4}

Now let's to study a scattering between two scalar fields and one photon (toy model - this is similar a scalar-electrodynamics) in phase space. This process is describes in FIG.5.
\vspace{1cm}
\begin{figure}[h]
\includegraphics[scale=0.8]{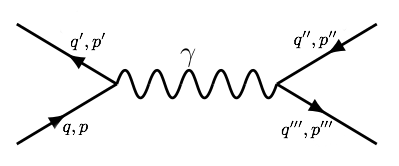}
\caption{The Feynman diagram for the process $\phi(q,p)+\phi^*(q',p')\rightarrow\phi(q'',p'')+\phi^*(q''',p''')$ with a photon in the intermediate state in phase space.}
\end{figure}

The transition amplitude for this process is given by

\begin{eqnarray}
-i \mathcal{M}=  \mathcal{V}^{\mu}_{\phi \phi^{\ast}A}\mathcal{G}_{\mu \nu}^{-1}\mathcal{V}^{\nu}_{\phi \phi^{\ast}A},\label{ta}
\end{eqnarray}
where $\mathcal{V}^{\mu}_{\phi \phi^{\ast}A}$ and $\mathcal{G}_{\mu \nu}^{-1}$ are given in eqs. (\ref{TV}) and (\ref{PP}), respectively. Using this result the cross section may be determined since it depends on the square of transition amplitude, i.e.,
\bea
\frac{d\sigma}{d\Omega}=\frac{1}{64\pi^2\,s}|{\cal M}|^2,\label{cs}
\eea
where $s=4E^2$ is the centre of mass energy. This application shown that is possible to study scattering processes in a scalar electrodynamics in phase space. An important note, the total scattering amplitude, eq. (\ref{ta}),  in Born approximation allow us to know the interaction potential between scalar field and photon in phase space, since the potential may be calculated by using the Fourier transform of the scattering amplitude \cite{MM1}, i.e.,
\bea
V(r)=\frac{1}{(2\pi)^3}\int{\cal M}\,e^{ip\cdot x}d^3p.
\eea
In addition, this result in phase space is to the first order expansion of the Moyal product. It worths to stress out that due to corrections of the processes in phase space there is a new contribution in the cross section which is very small once it is given in terms of $\hbar^2$.

\section{Conclusions}

In this article the scalar electrodynamics in phase space is studied. We obtained a propagator for the scalar field up to first order in the expansion of the Lagrangian density in terms of the Planck constant which appears in the Moyal product. In this context the photon propagator is the usual one. Thus the interaction between these fields are also corrected by such an expansion. That means that loop corrections are important in higher orders of Planck constant. We point out that such a procedure in phase space is much more complicated than in usual projected space. For instance one should note that for each process there are infinite orders of $\hbar$ in the Moyal product. Therefore loop diagrams have to be performed in first and second order of approximation of the Moyal product which forces us to include the field $A^\mu$ in such calculations. Higher order diagrams are necessary in order to analyze if the theory is finite or renormalisable, these important points will be considered elsewhere. We presented a transition amplitude for such a interaction process in phase space. The structure of scalar electrodynamics in phase space is entire analogous to the usual theory, maybe a coupling to gravitation could reveal a different behavior for the scalar field \cite{scalar06}. As a future work scalar electrodynamics in phase space can be thermalized by means Thermo Field Dynamics in order to study possible extra features.

\section*{Acknowledgments}

This work by A. F. S. is supported by CNPq project 308611/2017-9.


\end{document}